# Prequestioning Enhances Undergraduate Students' Learning in an Environmental Chemistry Course

Steven C. Pan[†*], Jia Yi Han[^], Fun Man Fung[^*]
[†]Department of Psychology, National University of Singapore, 9 Arts Link, Singapore 117570
[^]Department of Chemistry, National University of Singapore, 3 Science Drive 3, Singapore 117543
*Corresponding Author

**ABSTRACT**

*Prequestioning* is an instructional strategy that involves taking practice tests on to-be-learned information followed by studying the correct answers. Despite promising results in laboratory studies, it has rarely been examined in authentic educational settings. The current study investigated the pedagogical benefits of prequestioning as a learning intervention in an undergraduate environmental chemistry course. In each of 10 lecture sessions, the course lecturer administered four prequestions targeting concepts that were to be taught in the very next lecture session, then presented the correct answers. On assessments occurring during the next lecture session, there was evidence of a *prequestioning effect*—that is, better performance on questions targeting prequestioned concepts versus non-prequestioned concepts, in most cases. That benefit of prequestioning, which was relatively large (across all lecture sessions, an overall effect size of Cohen's $d$ = 2.04, $p$ < .001), highlights the utility of prequestioning as a promising approach for enhancing learning in undergraduate chemistry and similar courses.

**INTRODUCTION**

In the learning sciences, there is considerable evidence that engaging in practice testing *after* information has been learned or taught, which is formally known as *retrieval practice*, can improve long-term memory and support other beneficial learning outcomes.[1–4] More recently, researchers have begun exploring the possibility that engaging in practice testing *before* learning or instruction has occurred—a strategy formally known as *prequestioning* or *pretesting*—might also be helpful for learning.[5,6] Such work is an example of ongoing efforts to identity effective learning strategies that instructors can use to better prepare students for learning in their courses. These efforts have taken on greater importance since the COVID-19 pandemic.[4,7–9] In the research literature, there are already some indications that prequestioning can be useful for more than measuring what students already know about a given topic. Rather, such testing can improve student learning directly.[5,10]

Prequestioning involves (a) students attempting to answer practice test questions (i.e., prequestions) about information that they have not yet studied or received any instruction on, followed by (b) the opportunity to learn the correct answers to those questions. Typically, (a) involves one or more prequestions in short answer, cued recall, or multiple-choice question (MCQ) format, whereas (b) occurs through the provision of correct answer feedback (i.e., being shown the correct answers), receiving instructional materials to study (e.g., a textbook chapter or pre-recorded video), direct instruction (e.g., a live lecture), or some combination of those approaches.[5] Importantly, prequestioning targets information (i.e., course content) that students do not yet know, and consequently, their attempts to guess the answers are often incorrect.[10,11] Further, with prequestioning, (b) typically occurs shortly after students have finished answering the prequestions, thus providing an opportunity to learn the correct answers whilst memory for the prequestions is still strong. On a subsequent assessment of learning (which is typically referred to as a posttest), an advantage of prequestioning over no prior prequestioning (i.e., in a control condition that simply studied correct information) is often observed. That advantage is formally known as the *prequestioning effect* or the *pretesting effect*.[5,10,11]

Research questions

To date, most of the research on prequestioning has involved psychology laboratory studies.[5] That literature contrasts with the multiple classroom studies that have implemented retrieval practice in chemistry and other types of courses,[12–14] as well as other classroom studies have investigated non-test-based uses of pre-class activities.[15]  To date, very few investigations of prequestioning have occurred in authentic educational settings.  The current pilot study is among the first to investigate prequestioning in the domain of undergraduate chemistry education.  It was designed as a proof-of-concept for a prequestioning-based learning intervention that could be deployed on a regular basis in undergraduate chemistry or similar types of courses.

This study addressed the following research questions (RQ):

**Primary RQ:** What is the effectiveness of prequestioning on the learning of lecture content in an undergraduate course?

**Secondary RQ**: How do students react to the use of prequestioning—that is, what do their metacognitive judgments reveal about their beliefs and impressions of the instructional strategy?

**THEORETICAL FRAMEWORK**

A three-stage theoretical framework for pre-instruction testing effects proposed by Pan and Carpenter informed this study.[5] This framework consists of three stages—(1) Pre-instruction testing, (2) Learning opportunity, and (3) Posttest—that describe effects of prequestioning on learning outcomes. (1) involves the act of prequestioning itself, wherein a general mental process or state such as enhanced curiosity is triggered, memories involving a prequestion or information relating to the prequestion are formed, and/or other types of memories develop.  These events can influence (2), in which the correct answers are learned (e.g., via correct answer feedback, receiving instructional materials to study, etc.).  It is assumed that prequestioning usually enhances learning of the correct answers.  Finally, (3) entails the posttest, in which the learning processes that occurred during (1) and/or (2) support better performance relative to a non-prequestioning (i.e., control) condition, yielding a prequestioning effect.  Available empirical evidence informs much of the theorizing about the processes occurring in different stages of the framework, for instance studies addressing effects on attention and reading behavior following prequestioning.[6,16,17]

More broadly, prequestioning represents a candidate "desirable difficulty" – that is, a learning strategy that requires a greater amount of effort on the part of the learner and often yields a higher rate of errors, at least initially, but can improve learning over the long-term.[18]  Relative to other learning approaches such as reading learning objectives and studying correct information from the outset, prequestioning requires more effort to guess the answers to questions about unfamiliar content.  Doing so may be challenging or even uncomfortable for learners, particularly given the widespread aversion towards making mistakes (i.e., producing incorrect answers) among students.[19]  Whether the difficulties that prequestioning poses are in line with the desirable difficulties framework, however, requires determining whether it yields beneficial learning outcomes in authentic educational contexts.  Moreover, as is the case with other learning strategies,[13,20] students' metacognitive perceptions of prequestioning should be taken into account.[3,21,22]

**LITERATURE REVIEW**

Psychology laboratory studies have shown that prequestioning can, in a variety of instances, promote the long-term retention of knowledge and lead to other beneficial learning outcomes.  For a comprehensive accounting of such studies, the interested reader may consult any of several review articles.[5,23,24]  Multiple demonstrations of prequestioning effects in controlled laboratory settings exist.  As one example, Richland et al. investigated whether administering prequestions drawn from a text passage about achromatopsia prior to reading that passage would enhance learning relative to simply reading the passage only; across five experiments, a prequestioning effect was repeatedly observed on a posttest occurring immediately afterwards or up to one week later.[10]  In another example, Pan et al. investigated whether engaging in prequestioning prior to watching an online video lecture on the topic of statistics would increase attention, reduce mind wandering, and improve learning.  Across two experiments, they found that prequestioning did indeed yield all those beneficial outcomes.[16]

The few studies that have investigated prequestioning in authentic educational settings have also shown promising results.  For instance, Beckman administered prequestions prior to instructional units in an undergraduate aerospace course; doing so led to better subsequent high-stakes test performance relative to a condition that did not receive premcquestions.[25]  Soderstrom and Bjork had students in an undergraduate research methods course attempt a series of prequestions at the start of several lecture sessions; on a final exam conducted up to several weeks later, the students performed better on questions addressing previously prequestioned content or content that was closely related to it, versus questions addressing content that had not been prequestioned.[26] As a third example, Janelli and Lipnevich administered prequestions prior to five course modules in a Massive Open Online Course (MOOC) about climate change. On subsequent course exams, there was no benefit of doing so relative to a no-prequestioning condition among all students that engaged in prequestioning, many of whom dropped out of the course, but among the subset of students that completed the entire MOOC, a prequestioning effect was observed.[27]

In summary, although there are indications of prequestioning effects in real-world settings, the strategy has yet to be investigated extensively or at all in many widely-learned subject domains, including in the physical sciences. Further, effective implementations of prequestioning that can be deployed in a range of classroom contexts have yet to be fully developed. Addressing those research gaps was the driving motivation for the present study.

**RESEARCH METHODOLOGY**

The present study, which was conducted in an undergraduate chemistry course, investigated a novel prequestioning-based learning intervention that only required a limited amount of class time and was relatively simple to implement. Unlike prior investigations of prequestioning in classrooms, some of which required class time both before *and* after lecture sessions (to administer prequestions and assess their effects on learning, respectively), this study only used class time at the end of lecture sessions. The effects of this approach were measured on posttests conducted after a time span of at least two days (i.e., at the end of the next lecture session).

The study spanned 11 lecture sessions. At the end of each session, students received a worksheet packet containing practice test questions, which they were given 15 minutes to complete. That 15 minutes was used for postquestions that assessed learning as well as four prequestions (the prequestions would usually take about 5 minutes to complete). Afterwards, the course lecturer briefly presented the correct answers to each question before dismissing the class.

Setting and participants

Students enrolled in Environmental Chemistry (Course code: CM3261) at the National University of Singapore (NUS) in Semester 1 of the 2023-2024 academic year were recruited to participate (N=10). CM3261 is an elective chemistry course catered for undergraduate chemistry students in their third and fourth year of study.[12,28,29] As with most courses offered at NUS, CM3261 consists of 13 weeks of instruction. Each instructional week consists of two lecture sessions and one tutorial session, with the latter starting only on week 3 of the semester. The course is taught using a flipped classroom approach wherein students watch lecture videos made by the course lecturer prior to attending face-to-face lecture sessions.[30,31] During the lecture sessions, the lecturer will recap the contents taught in the videos before students work in their groups to complete an in-class assignment that is related to the overall topic for that session. They will then share their answers and opinions with their classmates before the lecturer provides the correct answers. Despite being a chemistry course, CM3261 adopts a multidisciplinary approach—that is, incorporating concepts from various subjects such as systems thinking,[32,33] planetary boundary framework, sustainability,[34–36] and UN SDGs[37–39]—to highlight the intricate relationships between human activities and the environment. A prior iteration of the course was the setting for a study of retrieval practice as implemented via a Telegram-based bot.[12] Overall, CM3261 is a content-heavy course which requires students to memorize and understand various terminologies and concepts that were not encountered in prior chemistry and chemistry-related courses.

Ethics approval

The study received ethics board approval (Departmental Ethics Review Committee (DERC), reference ID "Psych-DERC Reference Code: 2023-July-01) prior to its initiation. All students gave informed consent prior to participation.

Study preparation

Before the commencement of the study, the Principal Investigator (SP) obtained the list of students' names from the course lecturer (FM) that were enrolled to CM3261 and generated a unique identifier (CX, where X is a three-digit integer from 100 – 112) for each of the students in the list. To protect their privacy, this unique identifier was used by the students in lieu of putting their actual names on each worksheet packet.  Further, to alleviate student concerns that their responses might impact their evaluation on other aspects of the course, only the Principal Investigator had access to a list containing the association between each of the identifiers and the students' names.

Before the start of each face-to-face session, the lecturer developed a series of questions to be used in the relevant study materials for that session.  These materials are described next.

Materials

The prequestioning intervention and the posttest were conducted using 11 worksheet packets (henceforth, worksheets) administered across 11 lecture sessions.  The topics for CM3261 that were addressed in those sessions are detailed in Table 1 and in Figure 1.  Before the start of each session, the lecturer devised eight postquestions for topics to be covered in that session, four prequestions for topics to be covered in the next lecture session, as well as answers for each question.

The worksheets contained the following items in the following order:

- A cover page with instructions stating that the worksheets would survey student knowledge of the content that was covered that day as well as content to be covered in the next lecture.
- Eight postquestions. These questions involved MCQ format with six choices each.  For lecture sessions 2-10, out of these eight questions, four postquestions had already been posed in the previous worksheet (as prequestions), while the other four questions were new questions on the same topic (henceforth, these questions will be referred to as *tested postquestions* and *untested postquestions*, respectively).^*
- Four prequestions. These questions used an MCQ format identical to that of the postquestions and targeted concepts that would be taught in the next lecture.  These questions were described to the students as "advance questions" which previewed content that would be covered later.
- A set of confidence and knowledge rating survey questions addressing each of the prequestions. These questions required numerical responses.

   ^* Because it was the very first lecture session in the entire course, the eight postquestions in lecture session 1 were all entirely new.  Further, the final worksheet that was administered in session 11 consisted of only eight postquestions (there were no prequestions as that session was the final one in the course).

The prequestions (or "advance questions") served as the learning intervention, whereas the postquestions measured the outcome of the intervention (specifically, the presence or absence of a prequestioning effect were determined by comparing the students' performance on the tested versus untested postquestions, which addressed information that covered by the prequestions versus information that was not covered by the prequestions, respectively).

A PowerPoint slide deck for each lecture session was prepared to share the correct answers to all twelve MCQs with the students after their administration.  A sample of the worksheets, as well as the PowerPoint slide deck containing the answers, can be found in the Supporting Information.

At the end of the course, participants were also asked to fill out an exit survey which contained a series of questions about their experiences with the different types of questions that they encountered on the worksheets.

Procedure

On the first day of class, all students taking CM3261 were briefed by the Principal Investigator on the purpose of the study, as well as the types of data that would be collected.  Specifically, they were informed that a series of "survey worksheets" would be administered at the end of 11 lecture sessions to (a) measure their knowledge for information that had already been taught and would be subsequently taught, and to (b) provide students with an opportunity to practice answering test questions about course content.  They were also informed that although their participation in this study was voluntary, they would receive credit amounting to 5% of the overall course grade for completing the worksheets and associated survey questions (without regard for their specific responses to those items).  An alternative activity for students that declined to participate in the study (essay assignments) was offered.  All students consented to participate in the study and were each given a unique identifier code to use on all study materials.

Data collection commenced during week 3 of the semester.  A schematic of the procedure is presented in Figure 2.  During each face-to-face session, the lecturer distributed the worksheets to the students towards the end of the class.  The students were given up to 15 minutes to complete the worksheets, then submitted their work to the research assistant for marking.  They were reminded not to leave any questions blank and that it was appropriate to guess.  The remaining five minutes of the class were used by the lecturer to display the answers to all the questions via a projector screen at the front of the classroom.  This cycle continued until the last lecture session (on week 8 of the semester where the final topic of the course was taught).

Data scoring and analysis

All worksheets were marked by a research assistant using an answer key provided by the lecturer, with the scores recorded electronically.  Using two sets of worksheets as examples (see Table 2), the students' scores were recorded as follows: (1) scores for the four prequestions upon their initial administration; (2) scores for the four tested postquestions; (3) scores for the four untested postquestions.  The difference between (2) and (3) was then calculated for each student for every topic except for the postquestions on worksheet 1 (as it entailed the first topic of the course and there were no prequestions administered beforehand; thus, it is not possible to measure the prequestioning effect for that topic).

## RESULTS

In the following analyses, where statistical tests were performed, α was set at 0.05.

Analyses of prequestioning effects

To address our primary research question, we conducted one-sample $t$-tests comparing postquestion performance for tested (i.e., previously prequestioned) versus untested (i.e., not previously prequestioned) concepts separately for each lecture session.  This comparison determined whether prequestioning on a set of four concepts from a given lecture yielded better performance on postquestions administered during the next lecture session.  Results are shown in Table 3.  As shown in the table, there was a statistically significant difference (i.e., $p < .05$) in favor of the tested condition—that is, evidence of a prequestioning effect—in five lecture sessions (with effect sizes ranging from Cohen's $d$ of 0.57 to 1.62).  Moreover, the numerical mean difference in performance between tested and untested postquestions favored the tested condition in nine of ten lecture sessions.

A one-sample $t$-test on participant-level mean difference scores for tested versus untested postquestions across all lecture sessions showed evidence of a strong prequestioning effect, $t(9) = 6.46$, $p = .00012$, $d = 2.04$.  That analysis is consistent with examination of violin and box plots of postquestion performance as depicted in Figure 3.  In that figure, there were indications of a prequestioning benefit for all but one lecture session and a strong prequestioning effect overall.  The mean advantage of tested versus untested postquestions—that is, the magnitude of the prequestioning effect in terms of proportion correct—was 0.19 [95% confidence interval (C.I.) of 0.12, 0.25].  Together, these results suggest that students were better able to answer postquestions addressing concepts that had previously been prequestioned—in other words, prequestioning enhanced learning.

Analyses of learning from prequestions to postquestions

To determine the amount of learning that students achieved after attempting prequestions, we conducted one-sample $t$-tests comparing performance on prequestions versus tested postquestions.  That analysis revealed the extent to which students had improved in their understanding of tested concepts—that is, between the prequestions targeting those concepts and subsequent postquestions targeting the same concepts.  Results are shown in Table 4.  There was a statistically significant improvement in six lecture sessions (with effect sizes ranging from $d = 0.92$ to 2.67).  Moreover, the mean difference in performance between prequestions versus tested postquestions was positive in all cases.  A corresponding analysis on data combined across all lecture sessions also showed evidence of learning improvements, $t(9) = 6.68$, $p < .00001$, $d = 2.11$.  Together, these results indicate that students engaged in productive learning processes during the period between prequestioning and the corresponding tested postquestions at the end of the next lecture session.

Metacognitive ratings after answering prequestions and at the end of the course

To address our secondary research question, we examined students' metacognitive ratings.  Students' confidence in the correctness of answers to prequestions and corresponding knowledge level ratings are detailed in Table 5.  Across all lecture sessions, the mean confidence rating was 42.3% and the mean knowledge level rating was 37.8%. Thus, the students were not highly confident in their answers or in their knowledge when attempting prequestions.

Metacognitive exit survey data are detailed in Table 6. The exit survey posed a series of questions which contrasted the effects of retrieval practice (i.e., postquestions) and prequestioning. We performed Wilcoxon signed rank tests to determine whether there were any differences in ratings of the two practice test approaches with respect to learning and memory; there were no significant differences in all cases. With respect to students' comfort level with attempting prequestions versus postquestions, however, the ratings favored retrieval practice over prequestioning ($p$ = .00042). In essence, students were more comfortable with answering postquestions than prequestions.

## DISCUSSION

The present study investigated the utility of prequestioning as a learning intervention in a previously unexplored knowledge domain, namely undergraduate chemistry, and using a novel implementation of the strategy. It served as a proof-of-concept for a prequestioning approach that potentially could be used in a wide range of classroom contexts. Overall, prequestioning conferred substantial learning benefits, and did not yield deleterious effects, in an environmental chemistry course. Spending approximately 15 minutes of class time to engage in prequestioning was beneficial for learning in multiple cases. The overall effect size of the observed prequestioning effect, $d$ = 2.04, ranks as quite large among all known educational interventions.[40] Moreover, the observed 0.18 proportion correct prequestioning effect across lecture sessions translates to a gain of nearly two letter grades according to conventional grading scales. We conclude that the prequestioning benefits in this study were meaningful in academic terms—which reflects one of the first successful demonstrations of a prequestioning benefit in a physical sciences education context.

### Why did prequestioning enhance learning?

The observed prequestioning effects can be interpreted using the three-stage framework outlined at the outset of this manuscript.[5] In the first stage, at the moment where students attempted to answer the prequestions, their guessing performance was relatively poor (i.e., approximately two-thirds of the prequestions were answered incorrectly). That low performance, however, set the stage for productive learning processes to follow (possibly triggered by improved attention, awareness of knowledge gaps, or curiosity). In the second stage, students had the opportunity to learn the correct answers when they were presented immediately afterwards, from the pre-recorded lecture videos, or possibly during the brief recap at the start of the next lecture (although discussion of specific correct answers was not guaranteed at that point). Any of these processes potentially impacted performance in the third stage, wherein students were better able to remember previously prequestioned (i.e., tested) versus not-prequestioned (i.e., untested) information.

The course lecturer observed that students repeatedly requested to take the worksheets home with them for further study. Doing so was prohibited however due to the need to adhere to study procedures. Nevertheless, that anecdotal evidence is an indication of the motivational or metacognitive effects that the prequestions likely had on student learning. In addition, in the exit survey, 60% of the students stated that the used the prequestions to guide their studying behaviors, which suggests that they may have attempted to remember the questions and focused on finding or studying the answers afterwards.

The substantial improvement in performance from the prequestions to the tested postquestions, which was approximately 0.36 proportion correct across all lecture sessions, confirms that students successfully learned the correct answers to prequestioned concepts in most cases. Moreover, that result speaks to the power of prequestioning to trigger subsequent learning behaviors and enhance learning. Further, the contrast between the high rate of errors during the initial prequestioning stage and subsequent improved tested postquestion performance establishes prequestioning as a "desirable difficulty" – that is, it is more challenging to use than other learning strategies, but yields better learning over the long term.[18]

### Metacognitive and instructional considerations

One potential challenge for the use of prequestioning was that students were not very comfortable with the strategy. In fact, 40% of students reported being moderately or very uncomfortable with doing so. That result may stem from students' discomfort with making errors and incorrect guesses.[41] Relatedly, undergraduate students in different countries tend to endorse both effective and less effective study strategies as being useful for learning,[7] and may be less comfortable with more effective strategies. Students also tended to not be very confident in the accuracy of their answers to the prequestions (which generally aligned with their actual accuracy) and rated their knowledge of prequestioned material as incomplete (which also appeared to be an appropriate assessment). Together, these results suggest that students may have been perturbed by the experience of engaging in prequestioning and felt unsure about the underlying content. Relatedly, some metacognitive theories of learning (e.g., the discrepancy reduction model) suggest that students tend to study in an effort to reduce knowledge gaps.[42] Prequestioning made those gaps evident, which was probably beneficial for learning over the long term. Further, other research suggests that it is possible to teach students metacognitive knowledge and skills in chemistry courses,[20] which raises the possibility of incorporating discussions of prequestioning into such training.

It is also notable that students did not strongly differ in their ratings for prequestioning versus retrieval practice in terms of effects on memory, effects on understanding of questioned content, and effects on learning or motivation. Correspondingly, research in the learning sciences has found that both types of practice testing can be beneficial, at least for memory and understanding. The relative benefits of engaging in practice testing before or after information has been learned, however, have yet to be fully established (studies to date have had conflicting results).[6,43,44]

From a practical standpoint, the amount of preparation that the course lecturer had to undertake could be considered to be moderate. The lecturer had already engaged in a routine of preparing practice test questions for various courses; adapting such questions to serve as prequestions and/or postquestions was relatively simple. A standard worksheet design was developed and used throughout the course; beyond that, all that was required was to swap in the necessary questions for each lecture session and print out the worksheets prior to class (it is possible that future implementations of this prequestioning approach could use online quizzing platforms such as Canvas). In summary, the prequestioning intervention used in this study did not require an extraordinary amount of preparatory work.

Finally, it is notable that the prequestioning approach used in this study entailed asking prequestions at the end of lecture sessions, not before, and at least two days prior to subsequent lecture sessions covering prequestioned content. The fact that prequestioning effects were successfully observed is in line with recent indications in the research literature that the benefits of prequestioning persist over retention intervals of at least several days.[45] It is also in line with earlier findings that prequestioning with semantically rich materials can be beneficial even when prequestioning and subsequent learning activities do not occur back-to-back.[46]

Limitations and future directions

As a pilot study, the present study is limited in several ways. First, given the nature of the course and its typical enrollment patterns, the overall sample size was small. Nevertheless, that sample was sufficient to observe statistically significant prequestioning effects using a within-participants design. Future studies may be able to interrogate prequestioning effects on a much larger scale. A second limitation that also stemmed from practical considerations involved the inability to assign course materials to be prequestioned or not prequestioned in a counterbalanced manner. A future study that does so would enable ruling out any effects of the course materials on the observed prequestioning effects. Nevertheless, given that prequestioning effects repeatedly occurred across multiple lecture topics in the present study and given that the course lecturer did not deliberately choose more difficult or easier concepts to be prequestioned, we suspect that similar results would have been observed had full counterbalancing been used.

A related third limitation is that prequestioning did not benefit all tested concepts and lecture topics equally. A notable exception was the topic of Climate Change, with which there was no significant prequestioning effect (although see [27]). It remains to be determined why prequestioning was ineffective in that case, but as with other learning strategies, the benefits of such strategies may not always be universal nor of equal magnitude across different sets of materials.

A final limitation is that prequestioning was not compared against a potentially more competitive control condition such as the study of learning objectives.[47] Against such conditions, the magnitude of the observed prequestioning effects may have been reduced. Nevertheless, the present study establishes prequestioning as a potentially beneficial "value add" that can enhance learning beyond typical instructional methods.

Finally future research may investigate ways to address student discomfort with prequestioning, possibly by provide explanations, feedback, or via other means.

## CONCLUSIONS

In an environmental chemistry course, prequestioning (in the form of four multiple-choice prequestions presented at the end of lecture sessions, followed by brief correct answer feedback) enhanced performance on test questions administered two or more days later. That result indicates that devoting a few minutes per lecture session to engaging in prequestioning can be beneficial for student learning. We concluded that prequestioning is a promising approach for enhancing learning in chemistry and other physical science courses.

## AUTHOR INFORMATION

Corresponding Author

scp@nus.edu.sg
fun.man@nus.edu.sg

## ACKNOWLEDGMENTS

The authors thank the students in CM3261 for their participation. All authors have reviewed and approved this submission. This research was supported by a National University of Singapore Faculty of Arts & Social Sciences (FASS) grant awarded to S. C. Pan.

**Table 1. Summary of topics that were covered, as well as the questions that appeared in the worksheets that were distributed at the end of each lecture session.**

| Worksheet Number | Topic taught during lecture | *Postquestion* topic (8 questions) | *Prequestion* topic (4 questions) |
|---|---|---|---|
| 1 | Element Cycle | Element Cycle | Lithosphere |
| 2 | Lithosphere | Lithosphere | Hydrosphere |
| 3 | Hydrosphere | Hydrosphere | Waste Management |
| 4 | Waste Management | Waste Management | Environmental Toxicology |
| 5 | Environmental Toxicology | Environmental Toxicology | Air Pollution |
| 6 | Air Pollution | Air Pollution | Structure |
| 7 | Structure | Structure | Ozone Layer |
| 8 | Ozone Layer | Ozone Layer | Food Chain |
| 9 | Food Chain | Food Chain | Climate Change |
| 10 | Climate Change | Climate Change | Sampling |
| 11 | Sampling | Sampling | - |

**Table 2. An example of how students' worksheet data were organized after marking.**

| | | Prequestions | | Postquestions | | | |
|---|---|---|---|---|---|---|---|
| Student ID | Topic | Number of Questions Correct | Proportion of Correct Answers | Tested questions that were correct (Out of 4)[a] | Proportions of correct answers for (tested) postquestions | Untested questions that were correct (out of 4)[b] | Proportions of correct answers for (untested) postquestions |
| CX100 | | 2 | 0.5 | 1 | 0.25 | 2 | 0.5 |
| CX101 | | 2 | 0.5 | 3 | 0.75 | 1 | 0.25 |
| CX102 | | 2 | 0.5 | 3 | 0.75 | 3 | 0.75 |
| CX103 | | 2 | 0.5 | 2 | 0.5 | 0 | 0 |
| CX104 | Food Chain | 2 | 0.5 | 4 | 1 | 2 | 0.5 |
| CX105 | | 2 | 0.5 | 4 | 1 | 3 | 0.75 |
| CX106 | | 2 | 0.5 | 1 | 0.25 | 1 | 0.25 |
| CX107 | | 1 | 0.25 | 4 | 1 | 3 | 0.75 |
| CX108 | | 3 | 0.75 | 2 | 0.5 | 1 | 0.25 |
| CX112 | | 2 | 0.5 | 4 | 1 | 3 | 0.75 |

[a]Postquestions that had appeared in the prequestion section in the previous worksheet.
[b]Postquestions that had not appeared in the prequestion section in the previous worksheet

**Table 3. Summary of statistical analyses of postquestion performance, by lecture topic, comparing *tested* versus *untested* postquestions.**

| Worksheet Number | Topic taught during Lecture | t | df | Prequestioning effect mean difference (95% CI) | p-value | Effect size (d) |
|---|---|---|---|---|---|---|
| 2 | Lithosphere | 3.33 | 6 | 0.36 (0.095, 0.62) | .0157* | 1.26 |
| 3 | Hydrosphere | 1.00 | 6 | 0.071 (-0.1, 0.25) | .356 | 0.38 |
| 4 | Waste Management | 2.37 | 7 | 0.25 (0.00019, 0.5) | .0499* | 0.84 |
| 5 | Environmental Toxicology | 2.59 | 9 | 0.23 (0.028, 0.42) | .0294* | 0.82 |
| 6 | Air Pollution | 1.91 | 9 | 0.18 (-0.032, 0.38) | .0886 | 0.60 |
| 7 | Structure | 1.00 | 7 | 0.063 (-0.085, 0.21) | .351 | 0.35 |
| 8 | Ozone Layer | 4.58 | 7 | 0.38 (0.18, 0.57) | .00254** | 1.62 |
| 9 | Food Chain | 2.86 | 9 | 0.23 (0.047, 0.4) | .0187* | 0.91 |
| 10 | Climate Change | -0.41 | 8 | -0.056 (-0.37, 0.26) | .695 | 0.14 |
| 11 | Sampling | 1.81 | 9 | 0.20 (-0.05, 0.45) | .104 | 0.57 |

*Note.* * $p < .05$, ** $p < .01$; Test number 1 is not included because it did not have a comparison of tested versus untested questions. Prequestion effect mean difference refers to participant-level mean performance (proportion correct) on tested postquestions minus performance on untested postquestions.

**Table 4. Summary of statistical analyses of performance improvement from prequestions to *tested* postquestions, by lecture topic.**

| Worksheet Number | Topic addressed | t | df | Improvement mean difference (95% CI) | p-value | Effect size (d) |
|---|---|---|---|---|---|---|
| 2 | Lithosphere | 7.07 | 6 | 0.71 (0.47, 0.96) | 0.00040** | 2.67 |
| 3 | Hydrosphere | 6.48 | 6 | 0.50 (0.31, 0.69) | 0.00064** | 2.45 |
| 4 | Waste Management | 2.18 | 7 | 0.28 (-0.023, 0.59) | 0.065 | 0.77 |
| 5 | Environmental Toxicology | 6.86 | 9 | 0.48 (0.32, 0.63) | 0.000074** | 2.17 |
| 6 | Air Pollution | 2.91 | 9 | 0.28 (0.061, 0.49) | 0.017* | 0.92 |
| 7 | Structure | 0.75 | 7 | 0.094 (-0.20, 0.39) | 0.48 | 0.27 |
| 8 | Ozone Layer | 3.03 | 7 | 0.31 (0.069, 0.56) | 0.019* | 1.07 |
| 9 | Food Chain | 1.71 | 9 | 0.20 (-0.064, 0.46) | 0.12 | 0.54 |
| 10 | Climate Change | 2.10 | 8 | 0.22 (-0.022, 0.47) | 0.069 | 0.70 |
| 11 | Sampling | 4.88 | 8 | 0.44 (0.23, 0.65) | 0.0012** | 1.63 |

*Note.* * $p < .05$, ** $p < .01$

**Table 5. Confidence and knowledge rating results for prequestion responses by lecture topic.**

| Worksheet Number | Topic taught during Lecture | Mean confidence rating (95% CI) From 0 (none) to 100 (completely), how confident are you that the answer is correct? | Mean knowledge rating (95% CI) From 0 (none) to 100 (full), what is your knowledge level of the concepts/ process/info addressed by this question? |
|---|---|---|---|
| 1 | Element Cycle | 48.85 (38.40, 59.30) | 39.06 (30.49, 47.64) |
| 2 | Lithosphere | 39.06 (22.76, 55.36) | 40.00 (30.01, 49.99) |
| 3 | Hydrosphere | 45.28 (20.87, 69.69) | 39.79 (23.44, 56.14) |
| 4 | Waste Management | 60.38 (47.51, 73.24) | 52.25 (42.47, 62.03) |
| 5 | Environmental Toxicology | 38.92 (28.76, 49.07) | 35.88 (26.04, 45.71) |
| 6 | Air Pollution | 55.78 (29.48, 82.08) | 49.06 (25.29, 72.84) |
| 7 | Structure | 43.13 (27.98, 58.27) | 38.75 (23.93, 53.57) |
| 8 | Ozone Layer | 29.08 (17.90, 40.25) | 22.05 (9.52, 34.58) |
| 9 | Food Chain | 33.96 (23.45, 44.47) | 35.75 (22.85, 48.65) |
| 10 | Climate Change | 31.62 (18.92, 44.32) | 27.64 (14.11, 41.16) |

Due to sample sizes ≤10, confidence intervals were calculated using the *t*-distribution.

**Table 6. Summary of student exit survey results.**

| Question | Form of practice testing | Answer choices and results | | | | |
|---|---|---|---|---|---|---|
| | | Strongly improved | Moderately improved | Neither improved nor reduced | Moderately reduced | Severely reduced |
| Effects on memory for questioned content? | Prequestioning | 20% | 40% | 40% | 0% | 0% |
| | Retrieval practice | 20% | 70% | 10% | 0% | 0% |
| Effects on understanding of questioned content? | Prequestioning | 10% | 30% | 60% | 0% | 0% |
| | Retrieval practice | 20% | 50% | 30% | 0% | 0% |
| | | Very helpful | Moderately helpful | Neither helpful nor helpful | Moderately unhelpful | Very unhelpful |
| Effectiveness for helping you learn course content? | Prequestioning | 30% | 20% | 40% | 10% | 0% |
| | Retrieval practice | 40% | 50% | 0% | 10% | 0% |
| Effectiveness for identifying topics to study? | Prequestioning | 20% | 40% | 30% | 10% | 0% |
| | Retrieval practice | 40% | 40% | 20% | 0% | 0% |
| Effectiveness for motivating you to study? | Prequestioning | 40% | 40% | 20% | 0% | 0% |
| | Retrieval practice | 40% | 40% | 20% | 0% | 0% |
| | | Yes | No | | | |
| Did you use the questions to guide your studying? | Prequestioning | 60% | 40% | | | |
| | Retrieval practice | 60% | 40% | | | |
| | | Very comfortable | Moderately comfortable | Neither comfortable nor uncomfortable | Moderately uncomfortable | Very uncomfortable |
| How comfortable were you in answering…? | Prequestioning | 0% | 30% | 30% | 30% | 10% |
| | Retrieval practice | 50% | 40% | 0% | 10% | 0% |

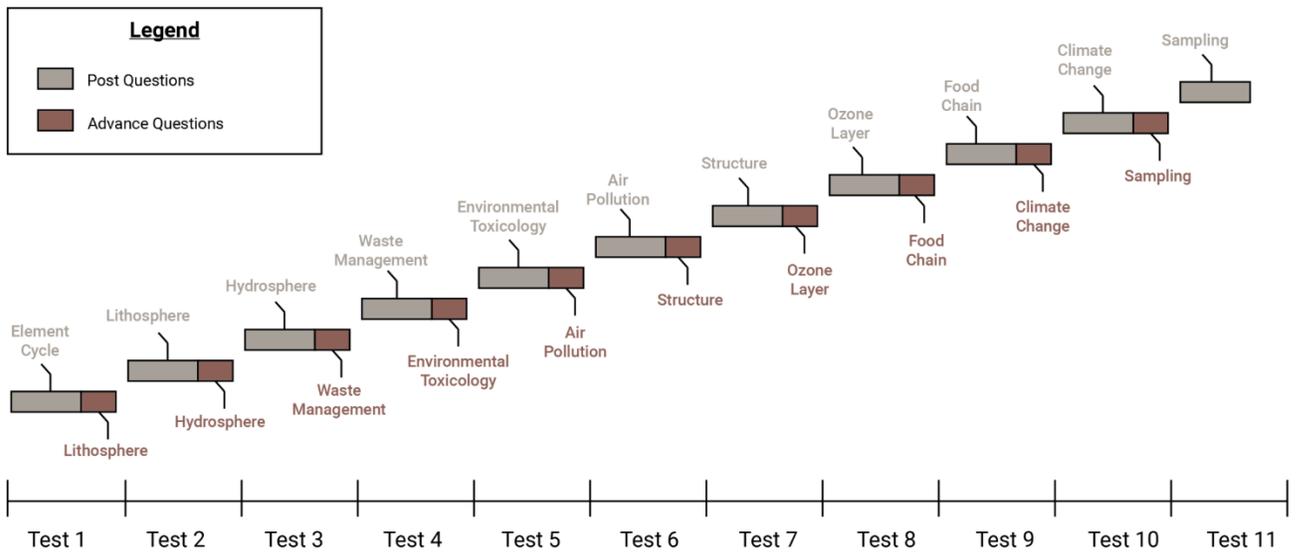

Figure 1. Timeline summarizing the topics tested for each of the 11 worksheets. The length of the postquestion bar signifies that there are more postquestions in that section as compared to the prequestions (i.e., "advance questions").

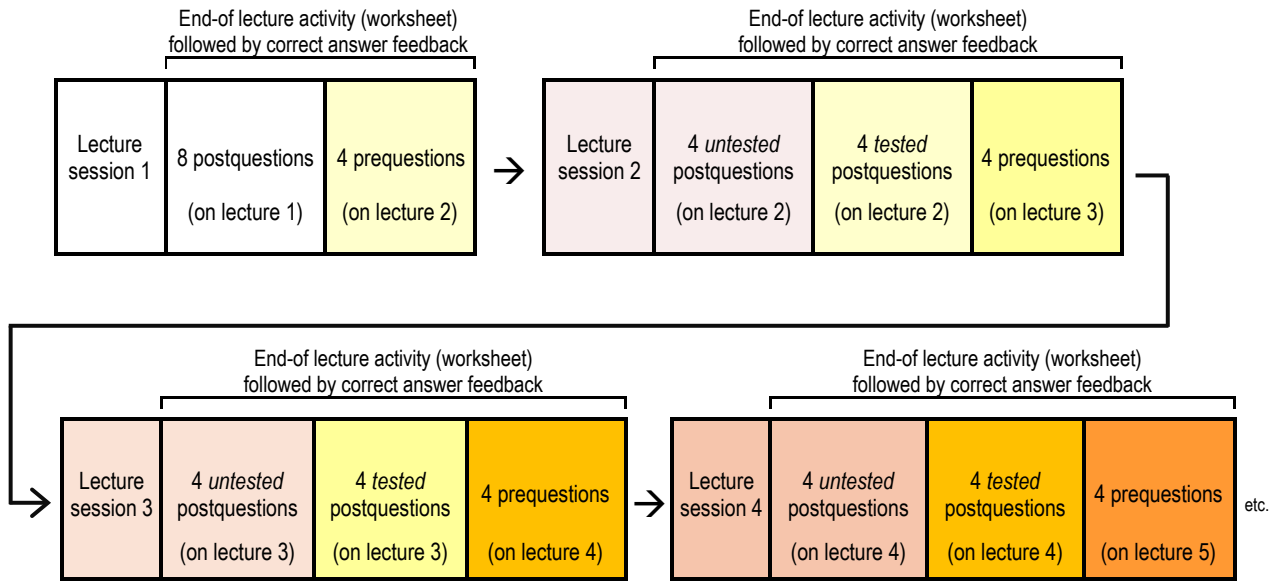

Figure 2. Experimental procedure showing the first four study cycles. At the end of each lecture session, students attempted 8 postquestions that assessed knowledge of content that was covered in that day's lecture, as well as 4 prequestions covering content to be taught in the next lecture. The course lecturer then presented the correct answers. For lecture sessions 2-11, the 8 postquestions were comprised of 4 *untested* (i.e., never-before-seen) postquestions as well as 4 *tested* postquestions (which were identical to the prequestions administered previously), randomly intermixed. (Note: because there was no opportunity to administer prequestions in any prior lecture session, the first lecture session lacked *tested* postquestions.)

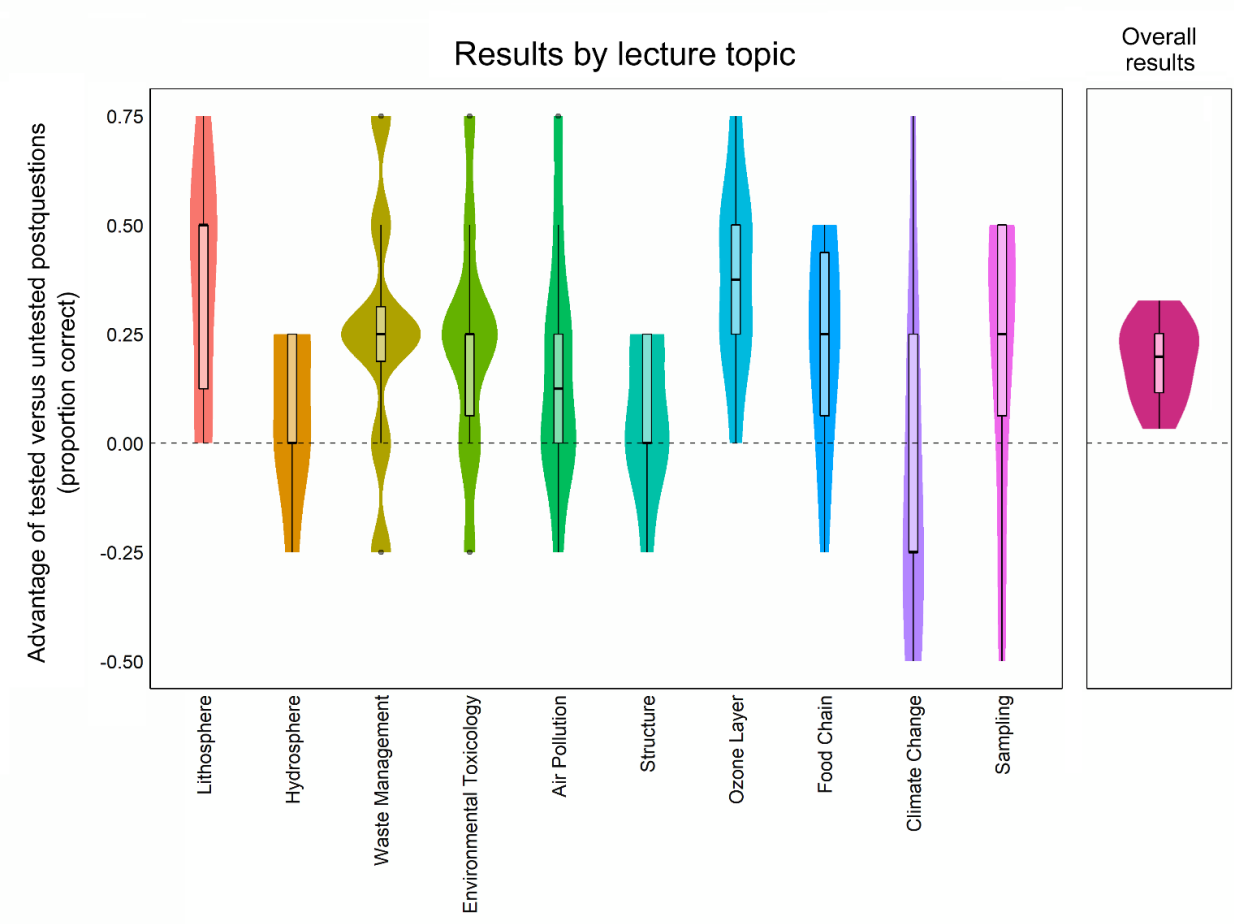

Figure 3. Benefits of prequestioning as indicated by a performance advantage for tested versus untested postquestions.  The left panel shows results arranged by lecture topic.  The right panel shows results collapsed across lecture topics.  The dotted line represents hypothetical lack of any difference between the tested and untested conditions.  See Table 3 for results of statistical analyses by topic.